\documentclass[prx,twocolumn,superscriptaddress]{revtex4-1}
\usepackage{amsmath, amsfonts, amssymb}
\usepackage{enumerate}
\usepackage{graphicx}
\usepackage{color}

\newcommand{\up}{\uparrow}
\newcommand{\dw}{\downarrow}
\newcommand{\e}{\mathrm{e}}
\newcommand{\mean}[1]{\langle #1 \rangle}
\newcommand{\mmean}[1]{\left\langle #1 \right\rangle}
\newcommand{\comm}[2]{[ #1, #2 ]}

\begin{document}


\title{Tunneling spectroscopy between one-dimensional helical conductors}

\author{Bernd Braunecker}
\affiliation{SUPA, School of Physics and Astronomy,
             University of St.\ Andrews, North Haugh, St.\ Andrews KY16 9SS, United Kingdom}

\author{Pascal Simon}
\affiliation{Laboratoire de Physique des Solides, CNRS, Univ.\ Paris-Sud,
University Paris-Saclay, 91405 Orsay Cedex, France}

\date{\today}


\begin{abstract}
We theoretically investigate the tunneling spectroscopy of a system of two parallel one-dimensional
helical conductors in the interacting, Luttinger liquid regime. We calculate the non-linear differential
conductance as a function of the voltage bias between the conductors and the orbital momentum shift
induced on tunneling electrons by an orthogonal magnetic field. We show that the conductance map
exhibits an interference pattern which is characteristic to the interacting helical liquid. This can
be contrasted with the different interference pattern from tunneling between regular Luttinger liquids
which is governed by the spin-charge separation of the elementary collective excitations.
\end{abstract}

\maketitle


\section{Introduction}

Topological phases and their accompanying exotic elementary excitations such as Majorana bound states
are currently a very active field of research, driven in part by the strong interest in applying
them for quantum information processing \cite{nayak:2008,alicea:2012,leijnse:2012,pachos:2012,beenakker:2013}.
A particular interest lies in one-dimensional (1D) topological systems due to an underlying prototype
model exhibiting Majorana edge states \cite{kitaev:2001,lieb:1961}, which has become a center of
attention since first proposals were made to realize it in 1D conductors placed in proximity of
superconductors \cite{fu:2008,sau:2010,oreg:2010,lutchyn:2010}.
The underlying condition is that the 1D conduction states are helical, which means they are spin filtered such that
electrons moving in opposite directions carry opposite spins. Such states can also have spintronic
applications, and it is thus a pertinent question whether there is a direct method to prove if a conductor
is or can become helical in some of its range of parameters.

This question is of course not new in this very developed field but we wish to address it in a
way that relies on the special many-body behavior characteristic of 1D conductors.
In 1D, due to the reduced dimensionality, electrons cannot avoid their neighbors
and necessarily all motion is collective.
The elementary excitations are indeed collective density wave modes, and their behavior
forms the basis of the Luttinger liquid (LL)
concept \cite{haldane:1981}. A hallmark of LL physics resulting from electron interactions is spin-charge
separation \cite{giamarchi:2004,gogolin:2004}.
The latter arises from the fact that the dynamics
of the low-energy collective spin and charge modes is governed by the spin susceptibility and the
charge compressibility, respectively. These are of different physical origin and thus are affected
differently by interactions. As a consequence, in a regular LL the spectral properties of the
spin and charge excitations decouple.

In a helical conductor, however, this decoupling breaks down because spin is pinned to the orbital motion.
Nonetheless the bound spin-charge fluctuations remain collective and they are described by the
helical Luttinger liquid (HLL) model
\cite{wu:2006,hou:2009,stroem:2009,teo:2009,tanaka:2009,liu:2011}. The latter has the
general behavior of a spinless LL but should  reveal its helical basis in the spin dependent
response functions.

In this paper we thus consider setups in which such response functions can be directly probed
and compared with existing experiments for regular LLs.
The experiments and modelling of tunneling between finite parallel quantum wires
performed in Refs. \cite{auslaender:2002,tserkovnyak:2002,tserkovnyak:2003,auslaender:2005,steinberg:2008}
provide one of the clearest evidences for LL behavior by complete tunneling spectroscopy,
in which the separated spin and charge spectra are directly visible. Importantly, due to the finite
wire size the conductance data reveals a characteristic interference pattern that is
caused by the spin-charge separation. Such double wires are therefore an ideal setup to identify unique characteristics
to distinguish between regular and helical LLs.

\begin{figure}
\centering
	\includegraphics[width=\columnwidth]{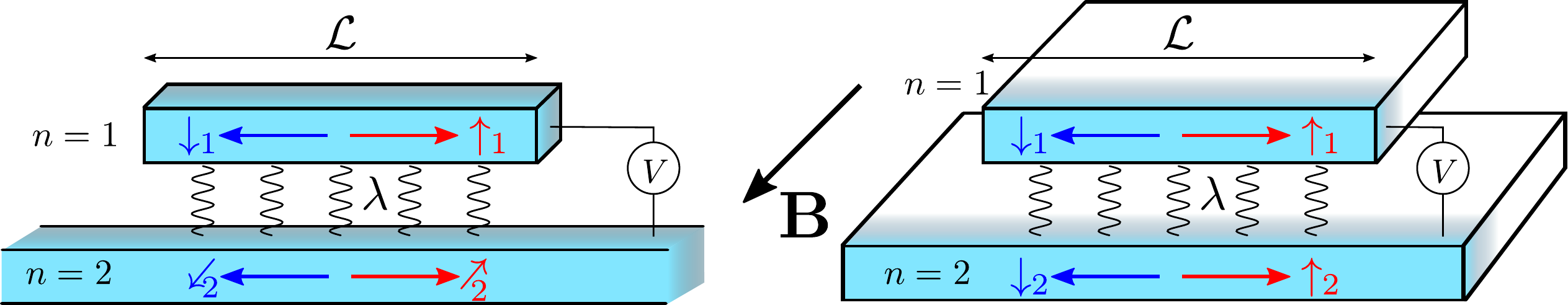}
	\caption{\label{fig:system}
		Sketch of two possible scenarios of tunneling between two quantum wires (left)
		or edge states of a bilayer system of a topological insulator (right).
		The 1D conducting modes (shaded areas) are helical, i.e.\ opposite
		spin directions are bound to opposite motion as indicated by the arrows.
		While the spin
		directions in a bilayer system are generally parallel, they can differ between individual
		quantum wires, represented by the rotated symbols $\up_2, \dw_2$ on the left.
		Tunneling between the conductors (wiggled lines) is of amplitude $\lambda$ and takes place over the length $\mathcal{L}$
		of the upper conductor.
		A magnetic field $\mathbf{B}$ is applied perpendicular to the tunneling plane and
		provides an orbital momentum shift that allows compensation of the momentum mismatch
		between the Fermi points of the upper and lower conductors (see Fig.\ \ref{fig:tunneling}),
		and a voltage $V$ is applied between the conductors.
	}
\end{figure}

HLLs can be obtained in at least three different ways that are amenable for a double conductor
setup: (a) most naturally on the edges of a quantum spin-Hall system such as a topological insulator
(see \cite{hasan:2010,qi:2011} for reviews);
(b) in quantum wires with strong spin-orbit interaction and an external magnetic
field \cite{streda:2003,pershin:2004,devillard:2005,zhang:2006,sanchez:2008,birkholz:2009,quay:2010,braunecker:2010};
and (c) through an ordering phase transition between conduction electrons and magnetic moments, such as nuclear spins,
embedded in the conductor \cite{braunecker:2009a,braunecker:2009b}.
For those systems the parallel conductor setup can, as sketched in Fig.\ \ref{fig:system}, be obtained either through
a bilayer system of topological insulators or through double wire setups.
For the double wire setup situation (c) is most appealing because the ordering is a transition to the
thermodynamic ground state and does not require fine tuning as for systems with strong spin-orbit interaction.
In addition the same sample used in Refs. \cite{auslaender:2002,tserkovnyak:2002,tserkovnyak:2003,auslaender:2005,steinberg:2008}
was investigated recently at very low temperatures \cite{scheller:2014,zumbuhl:2018} and provided signatures
for the ordering transition. Yet these signatures are based so far on the secondary effect of conductance reduction
with elimination of possible other explanations \cite{scheller:2014}
and on the analysis of the further temperature dependence of the conductance \cite{aseev:2017},
and an unambiguous detection of this phase, e.g.\ by tracing tunneling spectroscopy through the phase transition,
would be desirable.

In this work, we extend the analysis performed in  Refs. \cite{tserkovnyak:2002,tserkovnyak:2003} for the case of two HLLs. We show that the spin-to-momentum locking characteristic of an helical liquid drastically affects the transport properties between two  HLLs. We do so by
evaluating the non-linear conductance between two parallel HLLs as a function of the bias voltage $V$ and magnetic field $B$ and compare it the case of two regular LLs. We show that the absence of spin-charge decoupling in the HLLs changes substantially the interference pattern in the conductance map that can easily be visually distinguished from the standard interacting LL. Figures \ref{fig:conductance_helical} and
\ref{fig:conductance_helical_x} below provide our main results for this interference pattern.

The plan of the paper is as follows: in Sec. \ref{sec:parallel_conductors}, we present our model of two parallel helical conductors. In Sec. \ref{sec:tunneling}, we use the Keldysh formalism to express the current operator between the parallel 1D conductors in terms of the non-equilibrium Green functions of the HLLs and then evaluate this current. Sec. \ref{sec:maps} contains our main results which are summarized through the conductance maps between two HLLs as
a function of voltage $V$ and magnetic field $B$. Finally Sec. \ref{sec:conclusions} contains our conclusion and summary of results.
The estimation of complex multiple integrals is relegated to the appendix.

\section{Parallel helical conductors}
\label{sec:parallel_conductors}

We consider a system of two parallel 1D conductors subject to a perpendicular magnetic field, in a
situation where the length of one wire is confined by some potential to the length $\mathcal{L}$ while the
length of the other wire can be considered as infinite.
For regular LLs the tunneling transport between the conductors has been described and measured
in Refs. \cite{auslaender:2002,tserkovnyak:2002,tserkovnyak:2003,auslaender:2005,steinberg:2008}. It is the goal of this work to show that the
transport properties are significantly changed if the conductors are in the HLL regime.

Although there are intrinsic differences between the edge modes of topological insulators
and the helical modes in quantum wires that can be significant
for different response functions or higher energies  \cite{braunecker:2012} or in disordered
wires \cite{braunecker:2013}, these differences provide only minor corrections for the present
discussion and we will treat all the considered systems within the HLL formalism.

We consider the case in which both conductors are in the HLL phase in which
spin is locked with the direction of motion of the edge modes,
such that right moving modes $R$ with momenta near $k_F$ have the opposite spin to
left moving modes $L$ with momenta near $-k_F$. Generally the spin directions can vary from wire to
wire and do not have to be the same for the two considered conductors, although for bilayer systems
we expect them to be parallel. We will choose accordingly
a conductor dependent spin basis and denote the corresponding spin projections by $\sigma_n = \up_n, \dw_n$
for wire $n$, chosen such that $R$ movers have spin $\up_n$ and $L$ movers spin $\dw_n$,
but we emphasize again that generally $\sigma_1$ and $\sigma_2$ are not parallel.
If we linearize the spectrum near the Fermi points, the corresponding Hamiltonian
for the conductor $n=1,2$ reads
\begin{align}
	H_{n} =
	&- v_{Fn} \int dx
	\bigl(
		\psi_{n,R,\up_n}^\dagger i \partial_x \psi_{n,R,\up_n}
		-
		\psi_{n,L,\dw_n}^\dagger i \partial_x \psi_{n,L,\dw_n}
	\bigr)
\nonumber\\
	&+ \int dx \, dx' \ \mathcal{V}(x-x') \psi_n^\dagger(x) \psi_n^\dagger(x') \psi_n(x') \psi_n(x),
\end{align}
where $v_{Fn}$ is the Fermi velocity, $\psi_{n,r,\sigma_n}$ is the electron operator in conductor $n=1,2$,
with $r=L,R=-,+$ denoting the left and right moving modes, and $\sigma_n=\up_n,\dw_n$ the natural spin basis
of this conductor as described above.
The full electron operator is $\psi_n = \psi_{n,R,\up_n} + \psi_{n,L,\dw_n}$, and $\mathcal{V}(x-x')$ describes the
electron-electron interactions. For the helical conductor the operators $\psi_{n,R,\dw_n}$ and $\psi_{n,L,\up_n}$
are absent or can be neglected for the present analysis.

The step to bosonization is done by expressing the electron operators
in terms of boson fields $\phi_{n,r,\sigma_n}$ as
\begin{equation}
	\psi_{n,r,\sigma_n}(x)
	= \frac{\eta_{n,r,\sigma_n}}{\sqrt{2\pi a}}
	  e^{i r k_{Fn} x} e^{- i r \phi_{n,r,\sigma_n}(x)},
\end{equation}
where $k_{Fn}$ is the Fermi momentum, $a$ is a short distance cutoff, and
$\eta_{n,r,\sigma_n}$ are the Klein factors, or ladder operators, whose role is to
reduce the corresponding fermion number by 1.

Notice that $k_{Fn}$ depends directly on the applied voltage. If we assume as shown
in Fig.\ \ref{fig:system} a voltage $V$ applied on the the upper conductor ($n=1$)
then $k_{F1} = k_{F1}|_{V=0} + e V / v_F$ whereas $k_{F2}$ remains unchanged
(we will keep the notation $v_{F1} = v_F$ for simplicity).

It is then convenient to introduce an effective spinless description by setting
\begin{align}
	\phi_n   &= \frac{1}{\sqrt{2}}(\phi_{n,R,\up_n} + \phi_{n,L,\dw_n}),
\\
	\theta_n &= \frac{1}{\sqrt{2}}(\phi_{n,R,\up_n} - \phi_{n,L,\dw_n}),
\end{align}
in terms of which the bosonized Hamiltonian can be obtained in the standard way
\cite{wu:2006,hou:2009,stroem:2009,teo:2009,tanaka:2009,liu:2011}
and is given by
\begin{equation} \label{eq:H_HLL}
	H_n
	=
	\int \frac{dx}{2\pi} \, v_{n}
	\Bigl[ K_{n}^{-1} (\nabla \phi_{n})^2 +  K_{n} (\nabla \theta_{n})^2 \Bigr].
\end{equation}
The boson fields are chosen such that $-\nabla \phi_n/\pi$ measures fluctuations
of the electron density and $\nabla \theta_n/\pi$ is canonically conjugate to $\phi_n$.
We shall set $\hbar=1$ throughout this paper.
In this Hamiltonian $K_n$ is a parameter incorporating the interaction $\mathcal{V}$ such that
$K_n=1$ for a noninteracting system and $0 < K_n < 1$ for repulsive Coulomb interactions.
The renormalized velocity is given by $v_n = v_{Fn}/K_n$.

\begin{figure}
\centering
	\includegraphics[width=\columnwidth]{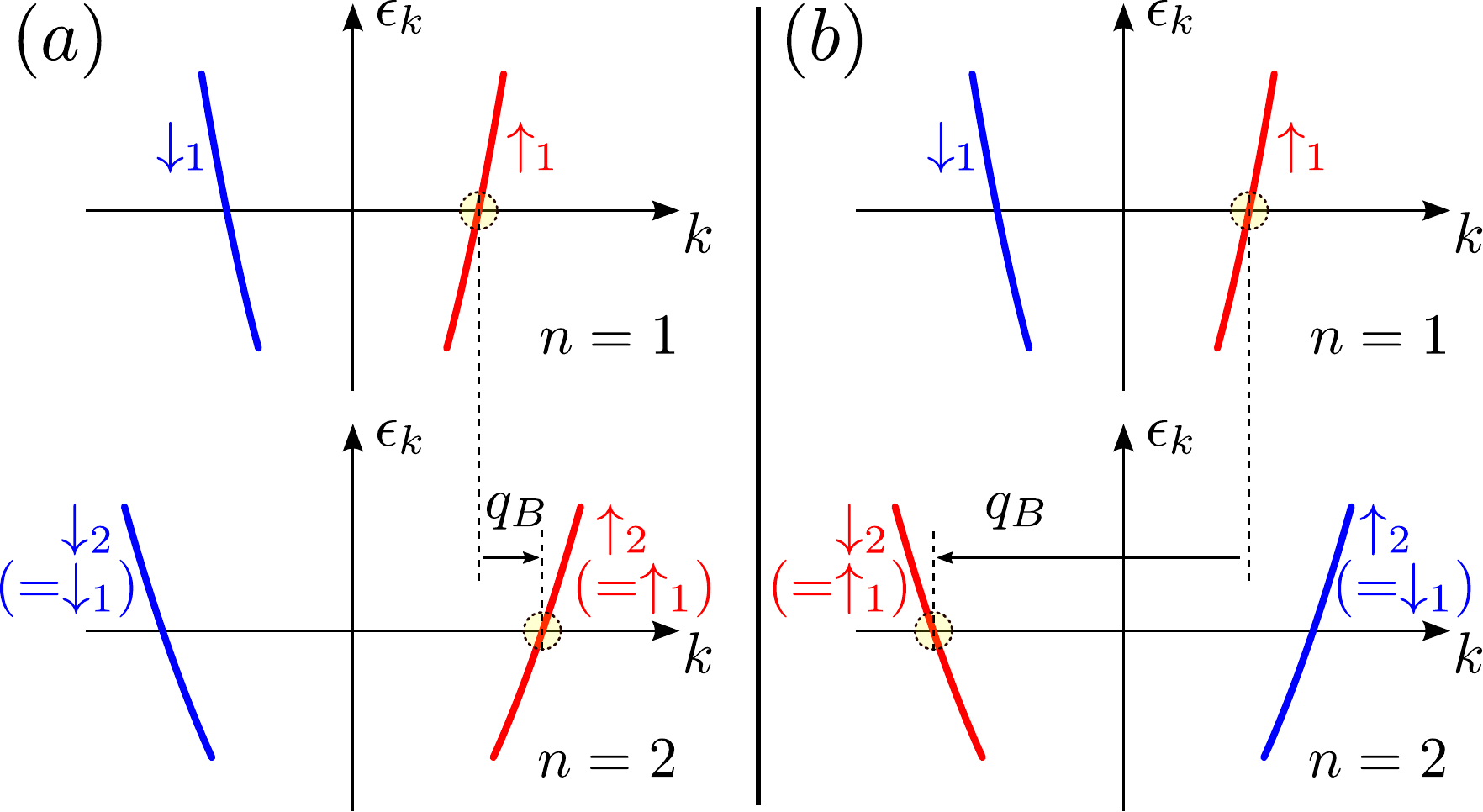}
	\caption{\label{fig:tunneling}
		Possible tunneling processes. The upper panels show the band structure of conductor
		$n=1$ and the lower panel of conductor $n=2$. The colors and arrows indicate the
		spin projections of the helical bands. The tunneling is spin conserving and the magnetic
		field must be tuned such that the momentum $q_B$ compensates for the mismatch of
		alignment of bands of the same spin projection. Case (a) corresponds to bands with
		identical spin structure but different densities (e.g.\ by different electrostatic
		environments or gating). Case (b) represents the special case of exactly opposite
		spin structures, requiring a large $q_B$ value going across both Fermi momenta.
		In the general case of non-collinear spin directions between the conductors both
		cases (a) and (b) are possible, weighted by the corresponding spin overlap matrix
		elements $\chi_{\sigma_1,\sigma'_2} = \langle \sigma_1 | \sigma'_2 \rangle$.
	}
\end{figure}

Tunneling between the conductors exhibits a few specialties. First, we consider that
one conductor, $n=1$, has a finite length $\mathcal{L}$. As highlighted in \cite{tserkovnyak:2002,tserkovnyak:2003}
the experimental current-voltage profile of regular LLs exhibits an asymmetry on the voltage sign. This asymmetry
can be reproduced by considering a soft confinement of the conductor instead of hard walls or
open boundary conditions.
Accordingly we model the confinement to length $\mathcal{L}$ by a soft envelope
function $\varphi(x)$ but assume that $\mathcal{L}$ is large enough such that Hamiltonian \eqref{eq:H_HLL}
remains valid.
Second, the orbital effect of the magnetic field $B$ applied perpendicularly to the conductors
leads to a phase shift $q_B x$ during tunneling, where $q_B = e B d$ \cite{steinberg:2008,lehur:2008}
with $e$ the electron charge and $d$ the distance between the conductors.
Finally, we should recall that the directions of the spin projections $\sigma_n = \up_n,\dw_n$
are not necessarily aligned between the conductors, and since the tunneling is spin preserving
(the tunneling distance is to small to allow for orbital precession) the tunneling amplitudes
are weighted by the spin overlap matrix elements
$\chi_{\sigma_1,\sigma'_2} = \langle \sigma_1 | \sigma'_2 \rangle$.
For a tunneling amplitude $\lambda$ the tunneling Hamiltonian then reads
\begin{align}
	H_T
	=
	\lambda
	\sum_{\sigma_1,\sigma_2'}
	&\chi_{\sigma_1,\sigma_2'}
	\int dx \, \Bigl[
		\varphi(x) \,
		e^{i q_B x - i e V t}  \psi_{2\sigma_2'}^\dagger(x) \psi_{1\sigma_1}(x)
\nonumber\\
		&+
		\varphi^*(x) \,
		e^{-i q_B x + i e V t} \psi_{1\sigma_1}^\dagger(x) \psi_{2\sigma_2'}(x)
	\Bigr],
\label{eq:H_T}
\end{align}
where $\psi_{n,\up_n} = \psi_{n,R,\up_n}$ and $\psi_{n,\dw_n}= \psi_{n,L,\dw_n}$ according to our
choice of the natural spin directions $\sigma_n$, and where we have furthermore introduced $V$, the
voltage drop between the conductors.
Due to energy and spin conservation only tunneling processes as shown
shown in Fig.\ \ref{fig:tunneling} are possible \cite{auslaender:2002,tserkovnyak:2002,tserkovnyak:2003,auslaender:2005,steinberg:2008,scheller:2014}.
If the spin projections of both conductors are parallel ($\chi_{\sigma_1,\sigma'_2} = \delta_{\sigma_1,\sigma'_2}$),
only tunneling between $L \to L$ and $R \to R$ movers are possible
and $q_B$ must be tuned as in Fig.\ \ref{fig:tunneling} (a) for the example $R \to R$, causing a matching of the
Fermi points as $k_{F2} = k_{F1} + q_B$.
On the other hand, Fig.\ \ref{fig:tunneling} (b) shows an example where only tunneling $L \to R$ and $R \to L$ is possible
($\chi_{\sigma_1,\sigma'_2} = \delta_{\sigma_1,-\sigma'_2}$),
requiring a large $q_B$ such that $-k_{F2} = k_{F1} + q_B$. In the general case of non-collinear spin projections (all $\chi_{\sigma_1,\sigma'_2} \neq 0$),
a tunnel
current flows for both settings of $q_B$ as in Fig.\ \ref{fig:tunneling} (a) and (b), albeit with further reduced amplitudes
by the $\chi_{\sigma_1,\sigma'_2}$.
This contrasts with the case of regular LLs in which tunneling between all $L$ and $R$ movers is possible with equal amplitudes,
and with the case of chiral LLs in which, for instance, $L$ movers are completely absent.

We should note that the tunneling occurs over the full length of the conductors, and for $\mathcal{L} \gg \pi /k_{Fn}$
the momentum resolution is fine enough such that tunneling only near the selected Fermi points is possible.


\section{Tunneling current between the 1D conductors}
\label{sec:tunneling}
This section is devoted to the  calculation of the tunneling current in terms of integrals involving the non-equilibrium Green functions and their estimation.

\subsection{Tunneling current}

The tunneling current operator is similar to Eq. \eqref{eq:H_T} and given by
\begin{align}
	I
	=
	-i \lambda
	\sum_{\sigma_1,\sigma_2'}
	&\chi_{\sigma_1,\sigma_2'}
	\int dx \, \Bigl[
		\varphi(x) \,
		e^{i q_B x - i e V t}  \psi_{2\sigma_2'}^\dagger(x) \psi_{1\sigma_1}(x)
\nonumber\\
		&-
		\varphi^*(x) \,
		e^{-i q_B x + i e V t} \psi_{1\sigma_1}^\dagger(x) \psi_{2\sigma_2'}(x)
	\Bigr].
\label{eq:I}
\end{align}
We will treat the current expectation value at order $\lambda^2$, which extends from
being perturbative to quantitatively exact for significant values of $q_B$ because
higher order process would not match the requirement of momentum conservation.
The standard Keldysh expression for the current
\begin{equation}
	\mean{I}
	= -i \int_{-\infty}^0 dt \, \mmean{\comm{I(0)}{H_T(t)}} ,
\end{equation}
is then rewritten in terms of Green's functions as
\begin{align}
	\mean{I}
	&= \lambda^2 \sum_{\sigma_1,\sigma_2'} |\chi_{\sigma_1,\sigma_2'}|^2
	\int dx dx' \, \varphi(x) \varphi^*(x')
	\int_{-\infty}^\infty dt \,
\nonumber\\
	&\times
	e^{i q_B (x-x') + i e V t}
		\Bigl[
			G_{1\sigma_1}^<(x-x',t) G_{2\sigma_2'}^>(x'-x,-t)
\nonumber\\
	&\qquad
			-
			G_{1\sigma_1}^>(x-x',t) G_{2\sigma_2'}^<(x'-x,-t)
		\Bigr],
\end{align}
with the greater and lesser Green's functions defined by
\begin{align}
	G_{n,\sigma_n}^>(x,t) &= -i \mean{\psi_{n\sigma_n}(x,t) \psi_{n\sigma_n}^\dagger(0,0)},
\\
	G_{n,\sigma_n}^<(x,t) &= +i \mean{\psi_{n\sigma_n}^\dagger(0,0)\psi_{n\sigma_n}(x,t)},
\end{align}
and again $\psi_{n,\up_n} = \psi_{n,R,\up_n}$ or $\psi_{n,\dw_n}= \psi_{n,L,\dw_n}$.
The Green's functions then become \cite{braunecker:2012}
\begin{align}
	&G^{>,<}_{n,\sigma_n}(x,t)
	=
	\frac{e^{i r k_{Fn} x}}{2\pi a}
	\nonumber\\
	&\times
	\left(\frac{a}{rx + v_{n}t_\mp}\right)^{\gamma_n}
	\left(\frac{a}{rx - v_{n}t_\mp}\right)^{\gamma_n+1},
\label{eq:G><}
\end{align}
where $r = +$ ($R$ mover) for $\sigma_n = \up_n$ and $r = -$ ($L$ mover) for $\sigma_n = \dw_n$.
Furthermore we have introduced the exponents $\gamma_n = \frac{1}{4}(K_{n} + K_{n}^{-1}-2)$,
and $t_\mp = t \mp i \eta$ for an infinitesimal $\eta>0$, with
$t_-$ corresponding to $G^>$ and $t_+$ to $G^<$.
We will restrain our analysis to zero temperature, which produces the correct description in the low temperature
regime whenever the voltage $V$ exceeds the thermal energy.

\subsection{Evaluation of the tunneling current}

For the calculation of the tunneling current we choose tunings of the $B$ field and voltage $V$ as shown
in Fig.\ \ref{fig:tunneling},
in which tunneling out of conductor $n=1$ occurs out of the $R$ moving modes with momenta near $+k_{F1}$
and into either the $R$ or $L$ movers in conductor $n=2$.

The spatial integration involves the function $\varphi(x)$ which is the sum of the different wave functions
that are confined by the finite length $\mathcal{L}$ of the conductor $n=1$, and its shape has been discussed in
Refs. \cite{governale:2000,boese:2001,tserkovnyak:2002,tserkovnyak:2003}. If $\mathcal{L} \gg \pi/k_{F1}$ the wave functions
in $\varphi(x)$ contain many nodes and consequently we can use the WKB approximation as well as the stationary
phase approximation to evaluate the spatial integration.
We thus write $\varphi(x) = C_L(x) e^{i s(x)} + C_R(x) e^{-i s(x)}$, where $s(x)$ adds a
phase within the wire but makes the function vanish outside the wire boundaries by turning complex,
and is given by
\begin{equation} \label{eq:s(x)}
	s(x) = k_{F1} \int_0^x dx' \, \sqrt{1-u(x')},
\end{equation}
where $u(x) = U(x)/E_{F1}$ for $U(x)$ the confining potential and $E_{F1}= v_1 k_{F1}/2$ the
Fermi energy of conductor $n=1$. We will choose $u(x) = (2x/\mathcal{L})^\beta$ for an even integer $\beta$
specializing, for instance, to $\beta = 8$ \cite{tserkovnyak:2002,tserkovnyak:2003}.

The amplitudes $C_{L,R}(x)$ are given by the standard WKB expressions, but since tunneling is restricted to
either $L$ or $R$ moving modes in conductor $n=1$ the WKB form must be restricted also to $L$ or $R$ moving
WKB solutions \cite{tserkovnyak:2002,tserkovnyak:2003}, for instance, for the $R$ moving case to $e^{- i s(x)}$.
This is not a harmless restriction but has the consequence that a real stationary phase
solution exists then only for positive voltages $V>0$ and is exponentially suppressed for $V<0$.
For an opposite magnetic field and tunneling out of the $L$ moving modes this dependence on the
sign of $V$ is inverted.

If we thus consider the case of tunneling out of $R$ moving modes as shown in Fig.\ \ref{fig:tunneling} then
only $C_R$ is nonzero.
We therefore choose $C_L=0$ and from the standard WKB approach we have
\begin{equation}
	C_R(x) = c [1-u(x)]^{-\frac{1}{4}}
\end{equation}
where the dimensionless constant $c\sim 1$ results from the
normalization condition $\int dx |\varphi(x)|^2 = \mathcal{L}$ (chosen such that $H_T$ remains
extensive without the need of rescaling $\lambda$). The precise value of $c$ depends on the exact shape
of the confining potential but is unimportant otherwise such that we can absorb it in the tunneling amplitudes
$\lambda$ and set $c=1$ henceforth.

Due to the highly oscillating phases we can evaluate the spatial integrals through the stationary phase
approximation, and a detailed discussion in found in App.\ \ref{app:stationary_phase}. In the limit of small
enough voltages such that $V \ll v_n/\mathcal{L}$ to guarantee tunneling into
the Fermi points separately, it is shown there that the spatial integration reduces to a saddle point
expression such that $x$ and $x'$ take only specific values $\pm x_0 \approx \mathcal{L}/2$
such that only the time integration
needs to be carried out. There are two contributions, one from $x-x'\approx 0$ and one from $|x-x'|\approx \mathcal{L}$.
To obtain a closed form for these integrals we assume that $v_1 = v_2 \equiv v$, since deviations from this
behavior do not have a significant influence and are usually small as the conductors need to be fabricated
on the same sample. We then write $\mean{I} = \mean{I}_{0} + \mean{I}_{\mathcal{L}}$, where $\mean{I}_{0}$ arises from
$x-x'\approx 0$ and $\mean{I}_{\mathcal{L}}$ from $|x-x'|\approx \mathcal{L}$.

The contribution $\mean{I}_0$ involves the Green's functions $G^{>,<}_{n\sigma_n}(x-x'=0,\pm t)$
and we need to evaluate
\begin{align}
	&\mean{I}_0
	\approx \frac{4\pi \lambda^2}{\beta k_{F1}} \sum_{\sigma_1,\sigma_2'} |\chi_{\sigma_1,\sigma_2'}|^2
	\Theta(Q_{\sigma_1,\sigma'_2})
	\int_{-\infty}^\infty dt \,
\nonumber\\
	&\times
	e^{i e V t}
		\Bigl[
			G_{1\sigma_1}^<(0,t) G_{2\sigma_2'}^>(0,-t)
			-
			G_{1\sigma_1}^>(0,t) G_{2\sigma_2'}^<(0,-t)
		\Bigr],
\end{align}
with $\Theta$ the unit step function and $Q$ the effective momentum transfer
\begin{align}
	Q_{\sigma_1,\sigma_2'}
	&= r_1 k_{F1}-r_2 k_{F2} + q_B
\nonumber\\
	&= (r_1 k_{F1}-r_2 k_{F2})|_{V=0} + q_B + r_1 eV/v_F,
\label{eq:Q}
\end{align}
where in the second line we have taken out explicitly the dependences on the magnetic field $B$ and
the voltage $V$.
The parameters $r_{1,2} = \pm$ for $R$ and $L$ moving modes, respectively, and are thus bound to the spin projections
as shown in Fig.\ \ref{fig:tunneling}.
Using Eq. \eqref{eq:G><} for the Green's functions we need to compute
\begin{align}
	&\int_{-\infty}^\infty dt \,
	e^{i e V t}
	G_{1\sigma_1}^{<,>}(0,t) G_{2\sigma_2'}^{>,<}(0,-t)
\nonumber\\
	&=
	\frac{(a/v)^{2\gamma}}{(2\pi a)^2}
	\int_{-\infty}^\infty dt \, e^{i e V t}
	\left(\frac{a}{+(t\mp i0)}\right)^{\gamma}
	\left(\frac{a}{-(t\mp i0)}\right)^{\gamma},
\end{align}
with $\gamma = \gamma_1+\gamma_2+1$ and where we have used $v_1=v_2=v$.
The phases of the power laws contain the crucial information for this
integration and evaluate to
\begin{align}
	\left(\frac{1}{t\mp i0}\right)^{\gamma}
	\left(\frac{1}{-t\pm i0}\right)^{\gamma}
	&= \begin{cases}
		t^{-2\gamma} e^{\pm i \pi (-\gamma)} &\text{for $t>0$}\\
		|t|^{-2\gamma} e^{\mp i \pi (-\gamma)} &\text{for $t<0$}
	\end{cases}
\nonumber\\
	&
	=
	|t|^{-2\gamma} e^{\mp i \pi \gamma \mathrm{sign}(t)}.
\label{eq:branches}
\end{align}
The time integration then turns into a standard Gamma function integral \cite{gradshteyn:1994},
\begin{equation}
	\int_0^\infty dt \, \sin(eVt) \, t^{-2\gamma}
	= \mathrm{sign}(eV) |eV|^{2\gamma-1}\frac{\pi \cos(\pi\gamma)}{ 2\Gamma(2\gamma) \sin(2\pi \gamma)},
\end{equation}
such that
\begin{align}
	\mean{I}_0
	&=
	\frac{4\pi \lambda^2 \mathcal{L}}{\beta k_{F1}} \sum_{\sigma_1,\sigma_2'} |\chi_{\sigma_1,\sigma_2'}|^2
	\Theta(Q_{\sigma_1,\sigma_2'})
\nonumber\\
	&\times
	\frac{(a/v)^{2\gamma}}{2\pi a^2 \Gamma(2\gamma)}
	\mathrm{sign}(eV) |eV|^{2\gamma-1}
\nonumber\\
	& =
	\mathrm{sign}(eV) |eV|^{2\gamma-1}
	\sum_{\sigma_1,\sigma_2'} \Theta(Q_{\sigma_1,\sigma_2'})
	\mathcal{T}_{\sigma_1,\sigma_2'},
\end{align}
where we have introduced the effective (dimensionful)
transmission coefficient
\begin{equation}
	\mathcal{T}_{\sigma_1,\sigma_2} = \frac{2 \lambda^2(a/v)^{2\gamma} \mathcal{L}}{\beta \Gamma(2\gamma)a^2 k_{F1}}
	|\chi_{\sigma_1,\sigma_2'}|^2.
\end{equation}
The second contribution $\mean{I}_{\mathcal{L}}$ maintains the spatial dependence on
$|x-x'| \approx \mathcal{L}$ and relies on the evaluation of
\begin{align}
	\int_{-\infty}^\infty dt \,
	&e^{i e V t}
		\Bigl[
			G_{1\sigma_1}^<(\pm \mathcal{L},t) G_{2\sigma_2'}^>(\mp \mathcal{L},-t)
\nonumber\\
	&
			-
			G_{1\sigma_1}^>(\pm \mathcal{L},t) G_{2\sigma_2'}^<(\mp \mathcal{L},-t)
		\Bigr].
\end{align}
An analysis of the branch structure of the power-laws in the Green's functions
following exactly the same evaluations as in Eq.\ \eqref{eq:branches} allows us to
rewrite the latter expression as
\begin{equation}
	4 \pi \sin(\pi\gamma)
	\frac{\e^{i (k_{F1}-r k_{F2})(\pm 2 x_0)}(a/v)^{2\gamma}}{(2\pi a)^2}
	\int_{\tau}^\infty \frac{\sin(eVt)}{(t^2-\tau^2)^\gamma},
\end{equation}
with $\tau = \mathcal{L}/v$. The remaining integral has a typical spurious divergence at $t \to \tau$
from the bosonization approach which needs to be regularized for $\gamma > 1$. This can be
straightforwardly done by an analytic continuation of the integration result for $\gamma < 1$ \cite{braunecker:2012}.
For the latter the calculation is reduced to a standard tabulated integral
\cite{gradshteyn:1994}, given by
\begin{align}
	&\int_{\tau}^\infty \frac{\sin(eVt)}{(t^2-\tau^2)^\gamma}
	= \mathrm{sign}(eV)
	2^{-\gamma-\frac{1}{2}} \sqrt{\pi}
\nonumber\\
	&\times
	|eV/\tau|^{\gamma-\frac{1}{2}}
	\Gamma(1-\gamma) J_{\gamma-\frac{1}{2}}(|eV \tau|),
\end{align}
with $J_{\gamma-\frac{1}{2}}$ the Bessel function.
Combining this result with all further factors and using the saddle point
expression of Eq.\ \eqref{eq:J_app} we obtain
\begin{align}
	&\mean{I}_{\mathcal{L}}
	=
	- \sum_{\sigma_1,\sigma_2'} \Theta(Q_{\sigma_1,\sigma_2'})
	\mathrm{sign}(eV) \left|\frac{v eV}{\mathcal{L}}\right|^{\gamma-\frac{1}{2}}
	\mathcal{T}_{\sigma_1,\sigma_2'}
\nonumber\\
	&\times
	C_\gamma
	\sin\bigl(Q_{\sigma_1,\sigma_2'}\mathcal{L} - 2 s(\mathcal{L}/2)\bigr)
	J_{\gamma-\frac{1}{2}}(|eV \mathcal{L}/v|),
\end{align}
where $s(\mathcal{L}/2)$ is given by Eq.\ \eqref{eq:s(x)}, $Q$ by Eq.\ \eqref{eq:Q},
and we have introduced the constant
\begin{equation}
	C_\gamma
	= \sqrt{\pi} \frac{\Gamma(2\gamma)}{\Gamma(\gamma)} 2^{\frac{1}{2}-\gamma}
	= 2^{\gamma-\frac{1}{2}} \Gamma\left(\gamma+\frac{1}{2}\right).
\end{equation}
The final result for the current, $\mean{I} = \mean{I}_0 + \mean{I}_{\mathcal{L}}$ can be
written as
\begin{align}
	&\mean{I}
	=
	\mathrm{sign}(eV)
	\sum_{\sigma_1,\sigma_2'}
	\Theta(Q_{\sigma_1,\sigma_2'}) \mathcal{T}_{\sigma_1,\sigma_2'}
	\Bigl[
		|eV|^{2\gamma-1}
\nonumber\\
		&
		-
		C_\gamma
		\left|\frac{v eV}{\mathcal{L}}\right|^{\gamma-\frac{1}{2}}
		\sin\bigl(Q_{\sigma_1,\sigma_2'}\mathcal{L} - 2 s(\mathcal{L}/2)\bigr)
		J_{\gamma-\frac{1}{2}}(|eV \mathcal{L}/v|)
	\Bigr].
\label{eq:I}
\end{align}
In the noninteracting limit $\gamma \to 1$ and $\mean{I} \propto V$ as  expected.

\section{Conductance maps}
\label{sec:maps}

\begin{figure}
\centering
	\includegraphics[width=\columnwidth]{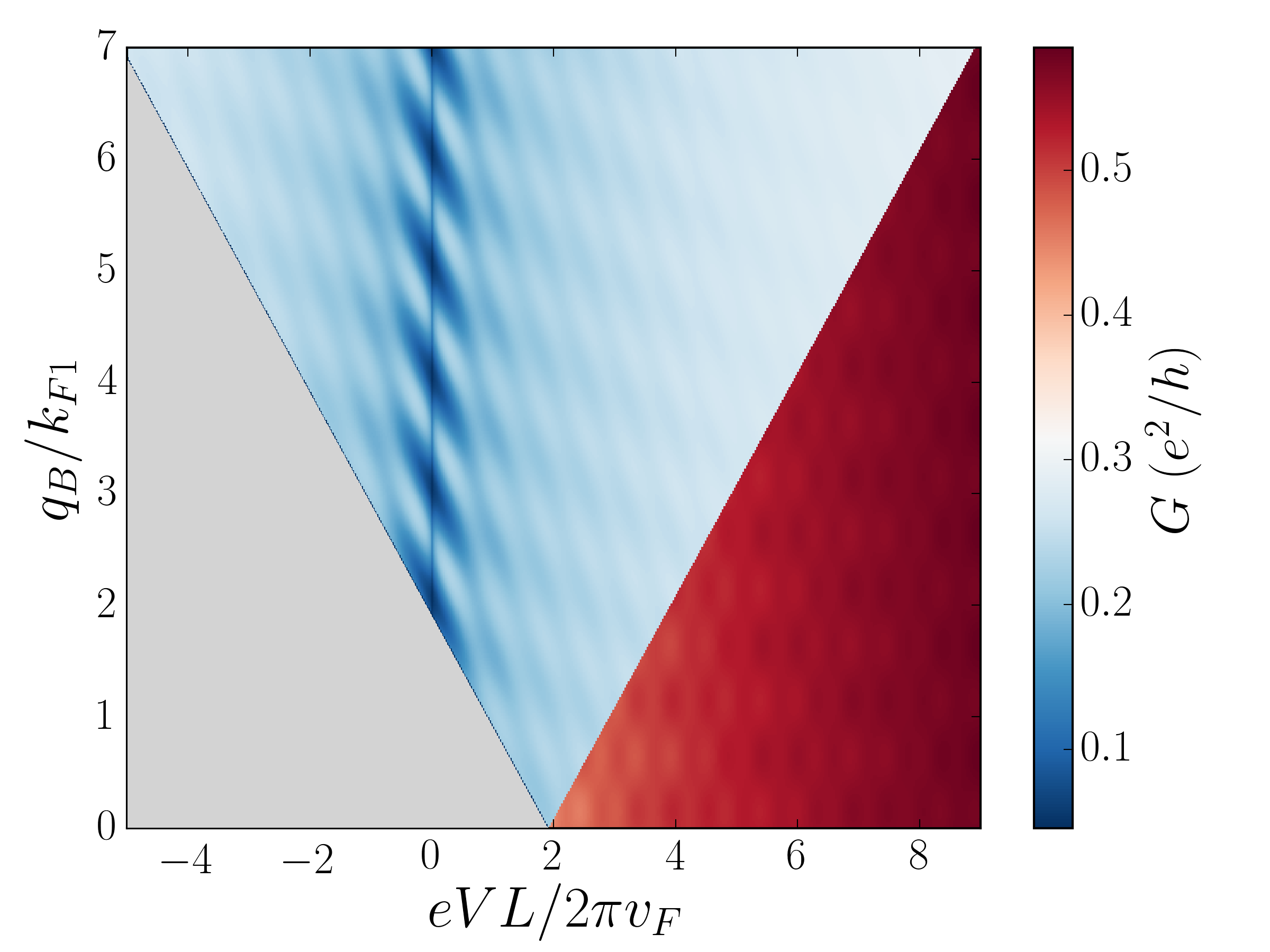}
	\caption{\label{fig:conductance_helical}
	Conductance map for tunneling between HLLs as a function of voltage $V$ and magnetic field $B$
	(through $q_B = e B d$). Assumed is $\sigma_1 = \sigma_2$ such that tunneling is only possible
	between the modes $R \to R$ and $L \to L$.
	The used parameters for all plots are $\mathcal{L} = 6\, \mu$m,
	$v_F = 2 \times 10^5$ m/s, $k_{F1} = 10^8$ m$^{-1}$ and $k_{F2} = 1.02 k_{F1}$ (at  $V=0$),
	$a = 5$ \AA, $\lambda = 1$ meV, $\beta=8$, and interaction parameters $K_1 = 0.8$, $K_2 = 0.6$.
	Since we set $v_1=v_2=v$ we choose $v = 2 v_F/(K_1+K_2)$ but notice that the precise
	interference beating pattern depends on this choice.
	The gray area marks $G=0$.
	}
\end{figure}
\begin{figure}
\centering
	\includegraphics[width=\columnwidth]{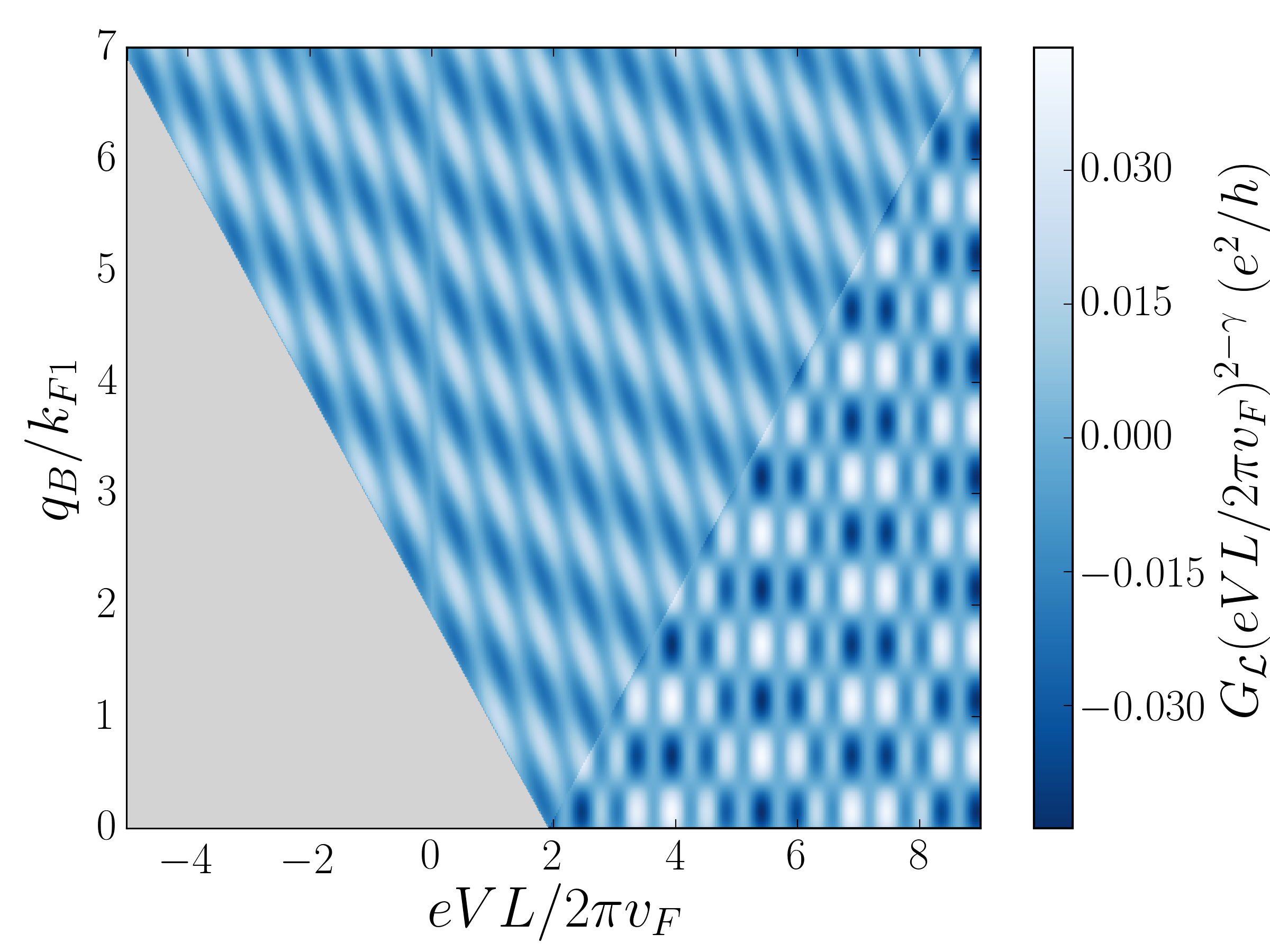}
	\caption{\label{fig:conductance_helical_x}
	Conductance part $G_\mathcal{L}$ as in Fig.\ \ref{fig:conductance_helical}, multiplied
	by $(eVL/2\pi v_F)^{2-\gamma}$ to enhance the interference effect by suppressing the
	power-law dependence on $V$. The interference pattern is the superposition of the
	sinusoidal and Bessel function oscillations of each tunneling process.
	}
\end{figure}
\begin{figure}
\centering
	\includegraphics[width=\columnwidth]{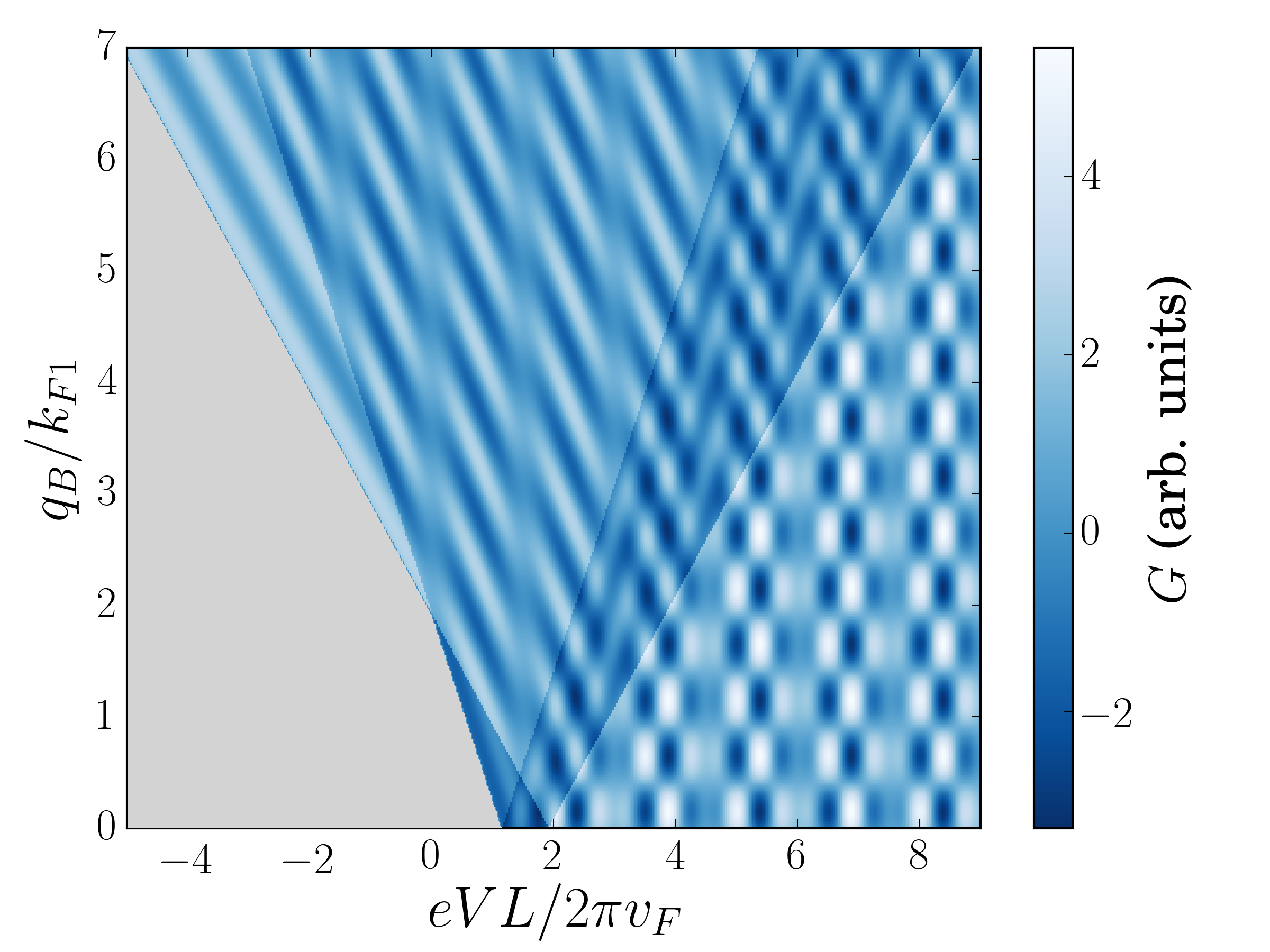}
	\caption{\label{fig:conductance_regular}
	Conductance part showing the interference pattern for tunneling between regular
	Luttinger liquids, as taken from Ref. \cite{tserkovnyak:2003}, for the same
	parameters as in Fig.\ \ref{fig:conductance_helical}, taking $K_{1,2}$ as the charge
	interaction strengths. The corresponding spin parameters are set to $K=1$ (representing
	a preserved spin SU(2) symmetry). The most notable difference to
	tunneling between HLLs are the vertical blurry interference fringes arising from the
	spin-charge separation. Notice that these fringes run continuously from the top to the bottom
	of the plot.
	}
\end{figure}

From the expression of the mean current in Eq.\ \eqref{eq:I} we obtain the differential conductance $G$ through $G = \partial \mean{I}/\partial V$.
If we let $G = G_0 + G_\mathcal{L}$ we have
\begin{equation}
	G_0
	=
	\frac{2\gamma-1}{e}
	\sum_{\sigma_1,\sigma_2'}
	\Theta(Q_{\sigma_1,\sigma_2'}) \mathcal{T}_{\sigma_1,\sigma_2'}
	|eV|^{2\gamma-2},
\end{equation}
where we disregard delta function contributions that would arise from differentiating the
step functions. The $G_\mathcal{L}$ contribution is dominated by differentiating the
$|v e V/\mathcal{L}|^{\gamma-\frac{1}{2}}$ term since the derivatives of the sine and
the Bessel function are by $|e V \mathcal{L}/v| \ll 1$ smaller. Hence
\begin{align}
	G_\mathcal{L}
	&=
	- \frac{\gamma-\frac{1}{2}}{|V|}
	\sum_{\sigma_1,\sigma_2'}
	\Theta(Q_{\sigma_1,\sigma_2'}) \mathcal{T}_{\sigma_1,\sigma_2'}
	C_\gamma
	\left|\frac{v eV}{\mathcal{L}}\right|^{\gamma-\frac{1}{2}}
\nonumber \\
	&\times
	\sin\bigl(Q_{\sigma_1,\sigma_2'}\mathcal{L} - 2 s(\mathcal{L}/2)\bigr)
	J_{\gamma-\frac{1}{2}}(|eV \mathcal{L}/v|).
\end{align}
In Fig.\ \ref{fig:conductance_helical} we show $G$ as a function
of applied voltage $V$ and magnetic field $B$ (through $q_B$) for the example
of Fig.\ \ref{fig:tunneling} (a) in which $\sigma_1=\sigma_2$. The
V-shaped structure arises from the step functions $\Theta(Q_{\sigma_1,\sigma_2'})$,
where $Q_{\up_1,\up_2} = (k_{F1}-k_{F2})|_{V=0} + eV/v_F + q_B$ corresponds to
$R \to R$ tunneling, and $Q_{\dw_1,\dw_2} = (-k_{F1}+k_{F2})|_{V=0} - eV/v_F + q_B$
to $L \to L$ tunneling. The gray area on the left side of the plot is where $G=0$.
In the central V-shaped area only $R \to R$ tunneling is possible. On the right
side of the V-shape the additional $L \to L$ tunneling leads to a considerable jump
in the magnitude of $G$. At $V=0$ there is a zero bias anomaly due to the characteristic
power-law divergence of a LL. But superposed are two types of oscillations.
One from the dependence on $\sin\bigl(Q_{\sigma_1,\sigma_2'}\mathcal{L} - 2 s(\mathcal{L}/2)\bigr)$,
providing oscillations parallel to the flanks of the V-shape, and one from the Bessel
function which is independent of $q_B$. Note that for the non-interacting case, $K_1=K_2=1$,
we recover a linear in $V$ behavior for the zero bias anomaly as it should be.
To visualize the interference pattern
in full detail we show in Fig.\ \ref{fig:conductance_helical_x} the part $G_\mathcal{L}$
only, multiplied by $V^{2-\gamma}$ to suppress further the power-law dependence.
In particular the superposition from $R\to R$ and $L\to L$ tunneling leads to a
characteristic checkerboard pattern.

In comparison we show in Fig.\ \ref{fig:conductance_regular} the corresponding
conductance map of a regular LL, using directly the results from Ref. \cite{tserkovnyak:2003}.
In the latter, the spin-charge separation causes the blurry vertical interference
pattern due to the difference of spin and charge velocities. This pattern is evidently
absent for the HLL. We observe furthermore in Fig.\ \ref{fig:conductance_regular}
a doubling of the V-shaped structure with different slopes for the spin and charge excitations,
arising from the different spin and charge velocities.
However, we believe  that this doubling could be due to an erroneous handling in Ref.\ \cite{tserkovnyak:2003} of the voltage dependence
in the step functions because the expansion in $V$ in Eq. \eqref{eq:Q}
relies on electrostatics and on the band structure and should thus involve $v_F$ throughout instead of
the interaction renormalized velocities. This would suppress the doubling of the V-structure.
But an unambiguous experimental resolution of this minor issue would be  welcome since an inspection
of the experimental data \cite{auslaender:2002,tserkovnyak:2002,tserkovnyak:2003} does not
allow us to reach a conclusion.

A further notable difference is the significantly
different checkerboard pattern on the right side of the V-shape.
For the HLL this different checkerboard pattern is a result of the interference of
the product of sinusoidal and Bessel function oscillations for the two tunneling
processes $R \to R$ and $L \to L$. But we should treat the precise pattern with care
since the Bessel function arises analytically from setting $v_1=v_2$ and hence the
result is strictly valid only for $K_1=K_2$, as we may expect it for bilayer systems
mostly. If the interactions differ, $K_1 \neq K_2$, we anticipate a further
beating effect, modulating the present pattern by a relative wavelength set by $(v_1-v_2)/v$.
We notice finally that in \cite{tserkovnyak:2003} the coupling between $x$ and $x'$ leading
to the Bessel function was neglected due to the stronger signatures by the spin-charge
separation, leading to the simpler interference pattern. For the HLLs we cannot
neglect this coupling and therefore predict the more involved patterns.

\begin{figure}
	\centering
	\includegraphics[width=\columnwidth]{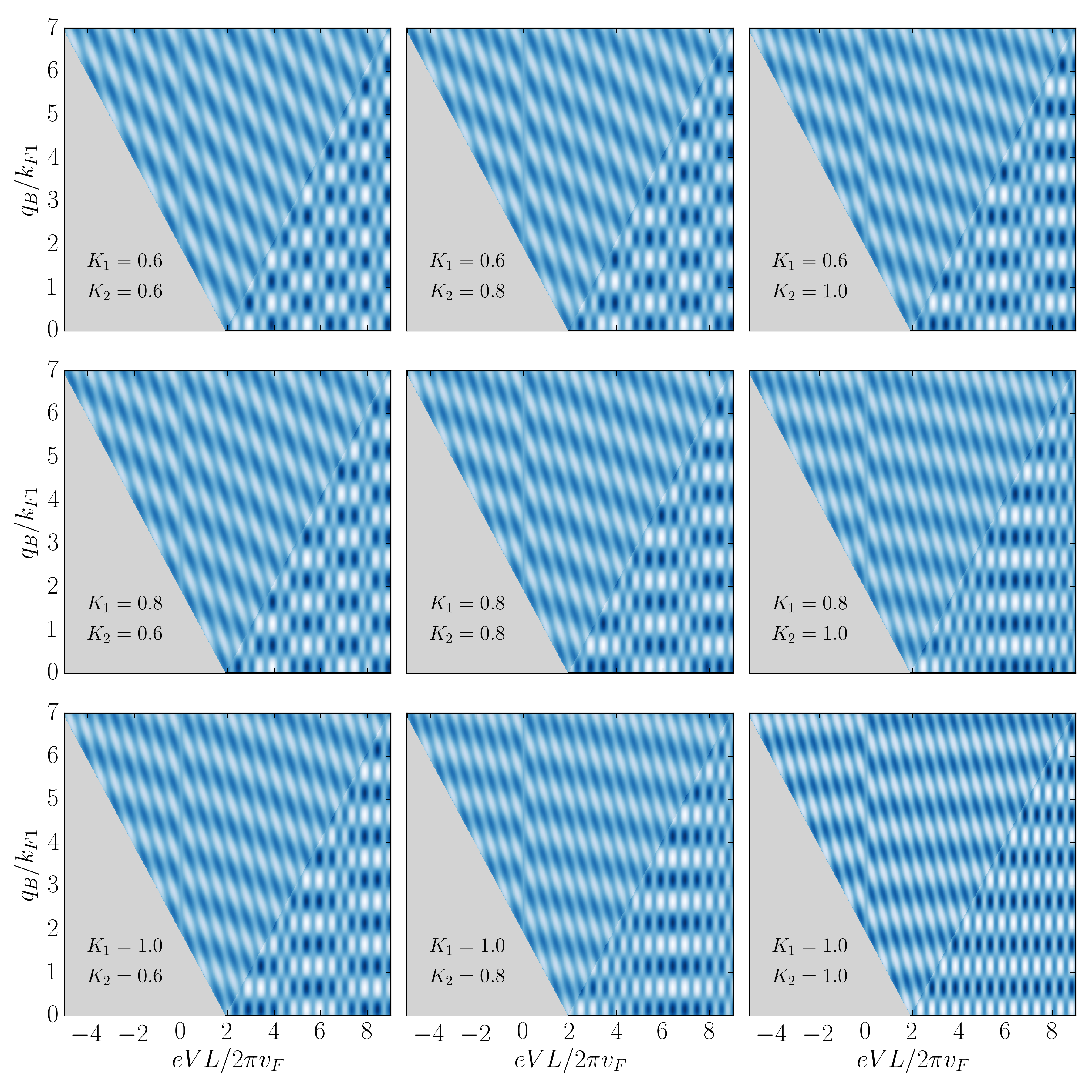}
	\caption{\label{fig:K_dep}
	Illustration of the change of the interference patterns
	as a function of the interaction strengths $K_1$ and $K_2$. Shown is
	$G_\mathcal{L} (eVL/2\pi v_F)^{2-\gamma}$ as in Fig.\ \ref{fig:conductance_helical_x}.
	The calculation is only exact for the plots on the diagonal where $K_1=K_2$, and
	for $K_1 \neq K_2$ we expect a further beating pattern from the velocity difference $v_1-v_2$.
	Notice that only in the noninteracting case $K_1=K_2=1$ the patterns exhibit
	horizontal uninterrupted beating fringes.
	}
\end{figure}
It is nonetheless worth to highlight that the interference patterns change with the
interaction strength, since the Bessel function depends on both the renormalized velocity
$v=v_F/K$ and the interaction dependent parameter $\gamma$.
In Fig.\ \ref{fig:K_dep} examples for a selection of $K_n$ values are shown.
Through the nonlinear dependence on the $K_n$ the changes of the interference patterns
are notable, and as the interactions are in part tunable with the electron density it
may be possible to track at least a part of this change and use
it to distinguish further the HLL from the regular LL case.

For tunneling processes $L \to R$ and $R \to L$, if again the spin overlaps permit,
the same results apply and
similar interference patterns are obtained. Yet to overcome now the large
momentum difference $k_{F1}+k_{F2}$ a large $q_B$ must be applied. Formally
one could also think of applying a large voltage but the present calculation
relies on the linearity of the bands in the considered energy ranges and it is
not possible to reliably extend the results to high voltages.


\section{Conclusions}
\label{sec:conclusions}

In this work we have calculated and analyzed the tunneling differential conductance between two
parallel helical 1D conductors as a function of the voltage bias $V$ and a magnetic field $B$ perpendicular
to the tunneling plane.
Our analysis applies either to two quantum wires where a low-energy helical behavior may emerge at low-enough
temperature or between two parallel edges of two 2D quantum spin Hall insulators (see Fig.\ \ref{fig:system}).
Our results are summarized mainly in Figs.\ \ref{fig:conductance_helical} and \ref{fig:conductance_helical_x}
where we have plotted the conductance map as a function of $B$ and $V$. Such maps are characterized by specific
interference patterns that result from the combination of the finite size confinement of the upper conductor
and the specific 1D physics. In the case of two standard Luttinger liquids the spin-charge
separation has the strongest imprint on the interference pattern in the form of blurry vertical stripes
(see Fig.\ \ref{fig:conductance_regular}), from whose oscillation period it is also possible to deduce
the interaction strength parameters $K$. For the helical conductors these stripes are absent and the
interference pattern has a very different visual aspect. We predict in particular a stronger role of
a futher oscillation pattern, expressed by a Bessel function with an
oscillation period that may still allow (for the case $K_1=K_2$) to extract the interaction strength.
But more significantly a detection of the change of the interference pattern can provide
a strong signature for entering a helical conduction state. This could especially lead to further
insight in the nature of the phase transition reported in Ref.\ \cite{scheller:2014}, where the
conclusions so far had to be based only on rather indirect signatures.


\begin{acknowledgments}
We would like to thank D. Zumb\"{u}hl for interesting discussions. BB acknowledges the hospitality
of the Universit\'{e} Paris-Sud and the Universidad Aut\'{o}noma de Madrid where part of this work was done.
Open data compliance: This work is theoretical and all plots are reproducible with the given formulas
and parameter values.
\end{acknowledgments}


\appendix

\section{Spatial integrals by stationary phase approximation}
\label{app:stationary_phase}

The different contributions to the tunneling current can generally be written in the form
\begin{equation}
	J = \int dx dx' \, \varphi(x) \varphi^*(x') e^{i Q (x-x')} g(|x-x'|),
\end{equation}
where $g(|x-x'|)$ contains the result from the time integration. This integral can be evaluated
through a stationary phase approximation. Since $x$ and $x'$ play the
same role in the integration the stationary phases are characterized either by $x=x'$ or $x=-x'$.
To collect the oscillating
parts of the integrand we notice that $g(|x-x'|)$ acquires through the time integration over $e^{i e V t}$
and the power-law dependences on $x-x' \pm v_n t$ the characteristic lengths $\lambda \sim v_n/V$.

At low voltages we have $\lambda \gg \mathcal{L}$, and the WKB envelope functions $\varphi$ dominate
the integration. We can then treat the $x$ and $x'$ integrations
individually. With $\varphi(x) \propto e^{-i s(x)}$ the stationary phase is obtained for
\begin{equation}
	i Q - i s'(x) = i Q - i k_{F1} \sqrt{1-u(x)} = 0,
\end{equation}
which can have a real solutions only for $Q>0$.
With the choice of $u(x) = (2 x/\mathcal{L})^\beta$ these real solutions are $x= \pm x_0$
with
\begin{equation} \label{eq:x0_app}
	x_0 = (\mathcal{L}/2)\left[ 1 - Q^2/k_{F1}^2 \right]^{1/\beta}.
\end{equation}
This stationary point is meaningful only for $|Q| < k_{F1}$, which is a natural condition to
consider for this modelling anyway, as $Q$ measures the momentum mismatch with the tunneling
between the two Fermi points of the conductors (cf. Fig.\ \ref{fig:tunneling}).
We recall also that this result relies on $\lambda \gg \mathcal{L}$ and thus does not allow
the extension to the case of an infinite conductor $\mathcal{L} \to \infty$. For the latter
a different approach such as in Ref. \cite{carpentier:2002} would be required.
Corrections to these saddle points from the coupling between $x$ and $x'$ through $g(|x-x'|)$
would be proportional to $g'(|x-x'|)/g(|x-x'|)$ and can be neglected in the present approximation.

Including Gaussian fluctuations in the stationary phase approximation
\begin{align}
	&\int dx \, e^{-i \Phi(x)}
	\approx \int dx \, e^{-i [ \Phi(x_0) + \frac{1}{2} \Phi''(x_0) (x-x_0)^2]}
\nonumber\\
	&=
	\sqrt{\frac{2\pi}{|\Phi''(x_0)|}}
		e^{-i \Phi(x_0)}
		e^{-i \frac{\pi}{4} \mathrm{sign}[\Phi''(x_0)]},
\end{align}
then leads to the solution
\begin{align}
	&J \approx \Theta(Q) \sum_{\nu,\nu' = \pm}
	\frac{2\pi e^{i Q (x_{\nu}-x_{\nu'})}}{\sqrt{|s''(x_\nu) s''(x_{\nu'})|}}
	\varphi(x_{\nu}) \varphi^*(x_{\nu'})
\nonumber\\
	&\times
	g(|x_\nu-x_{\nu'}|)
	e^{-i \frac{\pi}{4}\{\mathrm{sign}[s''(x_\nu)]-\mathrm{sign}[s''(x_{\nu'})]\}}
\end{align}
with $x_\nu = \nu x_0$.

For $u(x) = (2x/\mathcal{L})^\beta$ we have
$\mathrm{sign}[s''(x_\nu)] = \mathrm{sign}[u'(x_\nu)] = \mathrm{sign}(x_\nu) = \nu$
and $|s''(x_\nu)| = |s''(x_0)|$. This results in
\begin{align}
	&J
	\approx
	\Theta(Q) \sum_{\nu,\nu' = \pm}
	\frac{2\pi e^{i [Q x_0-\frac{\pi}{4}](\nu-\nu')}}{|s''(x_0)|}
	\varphi(x_{\nu}) \varphi^*(x_{\nu'})
\nonumber\\
	&\qquad\qquad\qquad\qquad\times
	g(x_\nu-x_{\nu'})
\nonumber\\
	&=
	\Theta(Q)
	\frac{4\pi |\varphi(x_0)|^2}{|s''(x_0)|}
	\bigl[
		g(0) - g(2x_0) \sin\bigl(2 Q x_0 - 2 s(x_0)\bigr)
	\bigr].
\end{align}
In this expression we should notice that $g(0)$ is independent on $\mathcal{L}$ and
hence larger than $g(2x_0)$ which through $x_0 \sim \mathcal{L}$ depends on $\mathcal{L}$
to some negative power.
But the $g(2x_0)$ term incorporates all interference effects and must not be neglected.

With $Q/k_{F1} = \sqrt{1-u(x_0)}$, $u(x_0) = (2x_0/\mathcal{L})^\beta$, and
$u'(x_0) = \beta u(x_0) / x_0$, we find
$|\varphi(x_0)|^2 = 1/\sqrt{1-u(x_0)} = k_{F1}/Q$ and
$|s''(x_0)| = k_{F1} u'(x_0)/2\sqrt{1-u(x_0)} = \beta k_{F1}^2 (2x_0/\mathcal{L})^\beta/2Q x_0$.
This allows us to write
\begin{equation}
	J \approx \Theta(Q)
	\frac{8\pi x_0 (\mathcal{L}/2x_0)^\beta}{\beta k_{F1}}
	\bigl[g(0) - g(2x_0) \sin\bigl(2 Q x_0 - 2 s(x_0)\bigr)\bigr].
\end{equation}
For the overall amplitude we can neglect the $Q$ dependence and
set $x_0 \approx \mathcal{L}/2$, which leads to the final approximation
\begin{equation} \label{eq:J_app}
	J \approx \Theta(Q)
	\frac{4\pi \mathcal{L}}{\beta k_{F1}}
	\bigl[g(0) - g(\mathcal{L}) \sin\bigl(Q \mathcal{L} - 2 s(\mathcal{L}/2)\bigr)\bigr],
\end{equation}
which is the result used in the main text.




\end{document}